\documentstyle[12pt,epsfig]{article}

\begin{document}


\mbox{ } \\[-1cm]
\mbox{ }\hfill DESY 99--198\\
\mbox{ }\hfill KIAS--P99111\\
\mbox{ }\hfill IFT--99/34\\
\mbox{ }\hfill hep--ph/0001175\\
\mbox{ }\hfill \today\\

\begin{center}
{\Large{\bf CHARGINO PAIR PRODUCTION \\[2mm]
            AT \boldmath{$e^+e^-$} COLLIDERS WITH POLARIZED BEAMS}}\\[1cm]
S.~Y.~Choi$^1$, M. Guchait$^2$, J.~Kalinowski$^3$ and P.~M.~Zerwas$^2$
\end{center}

\bigskip

\begin{enumerate}
\item[{}] $^1$ Korea Institute for Advanced Study, Seoul 130--012, Korea
\item[{}] $^2$ Deutsches Elektronen-Synchrotron DESY, D-22603 Hamburg, Germany
\item[{}] $^3$ Inst. Theor. Physics, Warsaw University, PL--00681 Warsaw, 
               Poland
\end{enumerate}
\bigskip

\begin{abstract}
The chargino $\tilde{\chi}^\pm_{1,2}$ system can be reconstructed completely
in $e^+e^-$ collisions. By measuring the total cross sections and the 
asymmetries with polarized beams in $e^+e^-\rightarrow\tilde{\chi}_i^+
\tilde{\chi}_j^- \, \,  [i,j=1,2]$, the chargino masses and the 
gaugino--higgsino mixing angles of these states can be determined very 
accurately.  If only the lightest charginos $\tilde{\chi}_1^\pm$
are kinematically accessible, transverse beam polarization is needed to 
determine the mixing angles unambiguously. From these observables the 
fundamental SUSY parameters can be derived: the SU(2) gaugino mass $M_2$, 
the modulus and the cosine of the CP--violating phase of the higgsino mass 
parameter $\mu$, and $\tan\beta = v_2/v_1$, the ratio of 
the vacuum expectation values of the two neutral Higgs doublet fields. 
[The remaining two--fold ambiguity of the phase can be resolved by
measuring the normal polarization of the charginos.]
\end{abstract}
%



\renewcommand{\thefootnote}{\alph{footnote}}
\newpage
\section*{1.~Introduction}

Charginos $\chi_{1,2}^\pm$ in supersymmetric theories are generally mixtures of
the spin--1/2 partners of the $W^{\pm}$ gauge bosons, $\tilde W^\pm$,
and of the charged Higgs bosons, $\tilde H^\pm$. The chargino mass
matrix \cite{R1} is given in the $(\tilde{W}^-,\tilde{H}^-)$ basis by
\begin{eqnarray}
{\cal M}_C=\left(\begin{array}{cc}
                M_2                &      \sqrt{2}m_W\cos\beta  \\
             \sqrt{2}m_W\sin\beta  &             \mu   
                  \end{array}\right)\
\label{eq:mass matrix}
\end{eqnarray}
The mixing matrix is built up by the fundamental supersymmetry (SUSY) 
parameters: the SU(2) gaugino mass $M_2$, the higgsino mass parameter $\mu$, 
and the ratio $\tan\beta=v_2/v_1$ of the vacuum expectation values of the two 
neutral Higgs fields which break the electroweak symmetry. 
By reparametrization of the fields, the phase $\Phi_\mu$ in
CP--noninvariant theories, may be attributed to $\mu$,
\begin{eqnarray}
\mu=|\mu|\,{\rm e}^{i\Phi_\mu}\ \ {\rm with}\ \ 0 \leq \Phi_\mu \leq 2\pi
\end{eqnarray}
while $M_2$ can be assumed real and positive; $\mu$ is real in CP--invariant
theories.\\

Since the chargino mass matrix
${\cal M}_C$ is not symmetric, two
different unitary matrices acting on the left-- and right--chiral
$(\tilde{W},\tilde{H})$ states are needed to diagonalize the matrix:
\begin{eqnarray}
U_{L,R}\left(\begin{array}{c}
             \tilde{W}^- \\
             \tilde{H}^-
             \end{array}\right)_{L,R} =
       \left(\begin{array}{c}
             \tilde{\chi}^-_1 \\
             \tilde{\chi}^-_2
             \end{array}\right)_{L,R}\,
\end{eqnarray}
The unitary matrices $U_L$ and $U_R$ can be parametrized in the
following way \cite{R11}:
\begin{eqnarray}
&& U_L=\left(\begin{array}{cc}
             \cos\phi_L & {\rm e}^{-i\beta_L}\sin\phi_L \\
            -{\rm e}^{i\beta_L}\sin\phi_L & \cos\phi_L
             \end{array}\right)\, \nonumber\\
&& U_R=\left(\begin{array}{cc}
             {\rm e}^{i\gamma_1} & 0 \\
             0 & {\rm e}^{i\gamma_2}
             \end{array}\right)
        \left(\begin{array}{cc}
             \cos\phi_R & {\rm e}^{-i\beta_R}\sin\phi_R \\
            -{\rm e}^{i\beta_R}\sin\phi_R & \cos\phi_R
             \end{array}\right)\,
\end{eqnarray}
The four phase angles $\{\beta_L,\beta_R,\gamma_1,\gamma_2\}$
are not independent but can be expressed in terms of  the invariant angle
$\Phi_\mu$. All four phase angles vanish in CP--invariant theories for which
$\Phi_\mu \rightarrow 0$ or $\pi$.\\

The mass eigenvalues $m^2_{\tilde{\chi}^\pm_{1,2}}$ and the rotation angles
$\phi_L$ and $\phi_R$ are determined by the fundamental SUSY parameters
$\{M_2,|\mu|, \cos\Phi_\mu,\tan\beta\}$;
\begin{eqnarray}
m^2_{\tilde{\chi}^\pm_{1,2}}=\frac{1}{2}\left[M^2_2+|\mu|^2+2m^2_W
                             \mp 4m^2_W \Delta\right]
\end{eqnarray}
and 
\begin{eqnarray}
&& \cos 2\phi_{L,R}=-\frac{M^2_2-|\mu|^2\mp 2m^2_W\cos 2\beta}{4m^2_W\Delta} 
   \nonumber\\
&& \sin 2\phi_{L,R}=-\frac{\sqrt{M^2_2+|\mu|^2\pm(M^2_2-|\mu|^2)\cos 2\beta
                +2M_2|\mu|\sin 2\beta \cos\Phi_\mu}}{2m_W\Delta}    
\end{eqnarray}
with $\Delta$ involving the phase $\Phi_\mu$
\begin{eqnarray}
\Delta=\sqrt{(M^2_2-|\mu|^2)^2
     +4m^2_W(M^2_2+|\mu|^2+2M_2|\mu|\sin 2\beta\cos\Phi_\mu)
     +4m^4_W\cos^2 2\beta}/4m^2_W
\label{eq:delta}
\end{eqnarray}
Conversely, the fundamental SUSY parameters $\{M_2,|\mu|,\tan\beta\}$
and the phase parameter $\cos\Phi_\mu$ can be constructed from the
chargino $\tilde{\chi}^\pm_{1,2}$ parameters: the masses
$m_{\tilde{\chi}^\pm_{1,2}}$ and the two mixing angles
$\phi_{L,R}$ of the left-- and right--chiral components of
the wave function (see Sect.4).\\

The two rotation angles $\phi_{L,R}$ and the phase angles
$\{\beta_L,\beta_R,\gamma_1,\gamma_2\}$ define the couplings
of the chargino--chargino--$Z$ vertices:
{\small
\begin{eqnarray}
&&{ }\hskip -5mm  \langle\tilde{\chi}^-_{1L}|Z|\tilde{\chi}^-_{1L}\rangle 
  = -\frac{g_W}{c_W} \left[s_W^2 - \frac{3}{4}-\frac{1}{4}
     \cos 2\phi_L\right] \,\qquad 
   \langle\tilde{\chi}^-_{1R}|Z|\tilde{\chi}^-_{1R}\rangle 
  = -\frac{g_W}{c_W} \left[s_W^2-\frac{3}{4}-\frac{1}{4}\cos 
    2\phi_R\right] \nonumber\\
&&{ }\hskip -5mm  \langle\tilde{\chi}^-_{1L}|Z|\tilde{\chi}^-_{2L}\rangle 
  = +\frac{g_W}{4c_W}\,{\rm e}^{-i\beta_L}\sin 2\phi_L\qquad 
  \qquad\quad\quad  \langle\tilde{\chi}^-_{1R}|Z|\tilde{\chi}^-_{2R}\rangle 
  = +\frac{g_W}{4c_W}\,{\rm e}^{-i(\beta_R-\gamma_1+\gamma_2)}
                        \sin 2\phi_R\nonumber\\
&&{ }\hskip -5mm  \langle\tilde{\chi}^-_{2L}|Z|\tilde{\chi}^-_{2L}\rangle 
  = -\frac{g_W}{c_W} \left[s_W^2 - \frac{3}{4}+\frac{1}{4}
     \cos 2\phi_L\right] \,\qquad
\langle\tilde{\chi}^-_{2R}|Z|\tilde{\chi}^-_{2R}\rangle 
 = -\frac{g_W}{c_W} \left[s_W^2-\frac{3}{4}+\frac{1}{4}\cos 
    2\phi_R\right]\nonumber 
\end{eqnarray}
}
and the electron--sneutrino--chargino vertices:
\begin{eqnarray}
&&{ }\hskip -5mm  \langle\tilde{\chi}^-_{1R}|\tilde{\nu}|e^-_L\rangle 
  = -g_Y\,{\rm e}^{i\gamma_1}\cos\phi_R \nonumber\\[1mm]
&&{ }\hskip -5mm  \langle\tilde{\chi}^-_{2R}|\tilde{\nu}|e^-_L\rangle 
  = +g_Y\,{\rm e}^{i(\beta_R+\gamma_2)}\sin\phi_R
\label{eq:vertex}
\end{eqnarray}
with $s_W^2 =1-c_W^2 \equiv \sin^2\theta_W$ denoting the electroweak mixing
angle. $g_W$ and $g_Y$ are the $e\nu W$ gauge coupling and the $e\tilde{\nu}
\tilde{W}$ Yukawa coupling, respectively. They are identical in supersymmetric
theories:
\begin{eqnarray}
g_Y=g_W=e/s_W
\end{eqnarray}
Since the coupling to the higgsino component, which is proportional
to the electron mass, can be neglected in the sneutrino vertex,
the sneutrino couples only to left--handed electrons.\\

Charginos are produced in $e^+e^-$ collisions, either in diagonal or in
mixed pairs \cite{R2}-\cite{7A}: 
\begin{eqnarray*}
e^+ e^- \ \rightarrow \ \tilde{\chi}^+_i \ \tilde{\chi}^-_j \ \ \ [i,j=1,2]
\,
\end{eqnarray*}
In the analysis of the chargino system the polarization of the electron 
and positron beams plays a central role. In addition to standard longitudinal 
(L/R) polarization, it turns out that transverse (T) polarization of the
beams will
be quite useful in the measurement of the chargino wave functions. The analysis 
will be carried out in two steps.\\
(i) In the first step we assume that
the collider energy will only be sufficient to generate the light 
$\tilde\chi^+_1 \tilde\chi^-_1$ pair. From the steep threshold behavior 
of the cross section, the mass of $\tilde\chi^\pm_1$ can be determined 
very accurately \cite{martyn}. From the size of the polarized cross sections 
in the continuum, the mixing parameters can be extracted. The transverse 
beam polarization is required in order to obtain a unique solution in general. 
Moreover, the mass $m_{\tilde\nu_e}$ of the sneutrino exchanged in the 
$t$--channel can be measured. \\
(ii) In the second step we assume the collider 
energy to be
large enough to produce the entire ensemble of diagonal and mixed chargino 
pairs $\tilde\chi_1^+ \tilde\chi_1^-$, $\tilde\chi_1^+ \tilde\chi_2^-$
and  $\tilde\chi_2^+\tilde\chi_2^-$.  In this case longitudinal beam 
polarization 
is sufficient to determine the mixing parameters unambiguously. 
Moreover, if the sneutrino mass is known from 
$\tilde\nu_e \bar{\tilde\nu_e}$ pair production, the 
threshold behavior and the continuum values of the polarized cross 
sections can be exploited to carry out a high--precision analysis of the 
masses $m_{\tilde\chi_{1,2}}$, the mixing parameters $\phi_{L,R}$ of the 
wave functions and the $e\tilde\nu_e\tilde W$ Yukawa coupling.\\

From these observables the underlying fundamental SUSY parameters,
$M_2,|\mu|$ and $\tan\beta$, can be extracted {\it unambiguously}; the
phase $\Phi_\mu$ can be determined up to a twofold ambiguity $\Phi_\mu
\leftrightarrow 2 \pi - \Phi_\mu$. This ambiguity can only be
resolved by measuring manifestly CP--noninvariant observables related to 
the normal polarization of the charginos, cf. Ref.~\cite{R11}.
To clarify the analytical structure, the reconstruction of the basic SUSY 
parameters presented here
is carried out at the tree level; the small higher--order corrections include 
parameters from other sectors of the MSSM demanding iterative higher--order
expansions in global analyses at the very end. \\
 
The analysis of the chargino sector is independent of the structure
of the neutralino sector\footnote{Many facets of the neutralino sector
have been discussed in the literature; for recent results see
Ref.~\cite{R12A} where mass relations are exploited, and Ref.~\cite{gmpsnu} 
where spin correlations have been considered.} which is potentially very 
complex in theories beyond the Minimal Supersymmetric Standard Model (MSSM). 
The structure of the chargino
sector, by contrast, is isomorphic to the form of the MSSM for a large
class of supersymmetric theories. Moreover, the analysis is based 
strictly on low--energy SUSY. Once these basic parameters are  
determined experimentally, they provide essential components in the 
reconstruction of the fundamental supersymmetric theory at the 
grand unification scale. 

\section*{2.~Chargino Production in $e^+e^-$ Collisions}

The production of chargino pairs at $e^+e^-$ colliders is
based on three mechanisms: $s$--channel $\gamma$ 
and $Z$ exchanges, and $t$--channel $\tilde{\nu}_e$ exchange, cf. Fig.1. 
The transition matrix element, after a Fierz transformation of the 
$\tilde{\nu}_e$--exchange amplitude,
\begin{eqnarray}
T[e^+e^-\rightarrow\tilde{\chi}^-_i\tilde{\chi}^+_j]
  =\frac{e^2}{s}Q_{\alpha\beta}
   \left[\bar{v}(e^+)\gamma_\mu P_\alpha  u(e^-)\right]
   \left[\bar{u}(\tilde{\chi}^-_i) \gamma^\mu P_\beta 
               v(\tilde{\chi}^+_j) \right]
\label{eq:production amplitude}
\end{eqnarray}
can be expressed in terms of four bilinear charges, defined by  
the chiralities $\alpha,\beta=L,R$ of the associated 
lepton and chargino currents. After introducing the following 
notation,
\begin{eqnarray}
&& \hskip -1.9cm 
   D_L=1+\frac{D_Z}{s_W^2 c_W^2}(s_W^2 -\frac{1}{2})(s_W^2-\frac{3}{4})\,
       \quad\qquad 
   F_L=\frac{D_Z}{4s_W^2 c_W^2}(s^2_W-\frac{1}{2})\ \nonumber\\
&& \hskip -1.9cm 
   D_R=1+\frac{D_Z}{c_W^2}(s_W^2-\frac{3}{4})\, \quad \quad \quad 
   \quad\qquad\qquad 
   F_R=\frac{D_Z}{4c_W^2}
\label{eq:DFLR}
\end{eqnarray}
and
\begin{eqnarray}
D'_L=D_L+\left(\frac{g_Y}{g_W}\right)^2\, 
              \frac{D_{\tilde{\nu}}}{4s^2_W}
   \qquad\qquad\hskip 1.3cm 
F'_L=\,F_L-\,\left(\frac{g_Y}{g_W}\right)^2\, \frac{D_{\tilde{\nu}}}{4s^2_W}
\end{eqnarray}
the four bilinear charges $Q_{\alpha\beta}$ are linear in the 
mixing parameters $\cos2\phi_{L,R}$ and $\sin2\phi_{L,R}$; for the diagonal 
$\tilde{\chi}^-_1\tilde{\chi}^+_1$, $\tilde{\chi}^-_2\tilde{\chi}^+_2$
modes and the mixed mode $\tilde{\chi}^-_1\tilde{\chi}^+_2$:\\
\begin{eqnarray}
\{11\}/\{22\} : &&{ } \hskip -5mm Q_{LL}=D_L\mp F_L\cos 2\phi_L\, 
                 \hskip 0.41cm \quad \qquad
                 Q_{RL}=D_R\mp F_R\cos 2\phi_L\, \nonumber \\
            &&{ } \hskip -5mm Q_{LR}=D'_L\mp F'_L \cos 2\phi_R\, 
	            \quad \qquad\hskip 0.36cm 
                 Q_{RR}=D_R\mp F_R\cos 2\phi_R \\
            &&                                \nonumber\\
{ } 
\{12\}/\{21\}:  &&{ }\hskip -5mm Q_{LL}=F_L\,{\rm e}^{\mp i\beta_L}\sin 2\phi_L\, 
                 \quad \quad \quad \quad\,\hskip 1mm
                 Q_{RL}=F_R\,{\rm e}^{\mp i\beta_L}\sin 2\phi_L\, \nonumber\\ 
            && { }\hskip  -5mm Q_{LR}=F'_L\,{\rm e}^{\mp i(\beta_R-\gamma_1
	        +\gamma_2)}\sin 2\phi_R\, \quad
                 Q_{RR}=F_R\,{\rm e}^{\mp i(\beta_R-\gamma_1
		 +\gamma_2)}\sin 2\phi_R\,
\label{eq:[12]}
\end{eqnarray}
The first index in $Q_{\alpha \beta}$
refers to the chirality of the $e^\pm$ current, the second index to the 
chirality of the $\tilde{\chi}_{i,j}^\pm$ current. The $\tilde{\nu}$ 
exchange affects only the $LR$ chirality charge $Q_{LR}$ 
while all other amplitudes 
are built up by $\gamma$ and $Z$ exchanges only. The first term in 
$D_{L,R}$ is generated by the $\gamma$ exchange;
$D_Z=s/(s-m^2_Z+im_Z\Gamma_Z)$ denotes the $Z$ propagator and
$D_{\tilde{\nu}} = s/(t- m_{\tilde{\nu}}^2)$ the $\tilde{\nu}$ propagator
with momentum transfer $t$. The non--zero $Z$ width can in general be 
neglected for the energies considered 
in the present analysis so that the charges are rendered complex in 
the Born approximation only through the 
CP--noninvariant phase.\\   

For the sake of convenience we introduce eight quartic charges for each of 
the production processes of the diagonal and mixed chargino pairs,
respectively. These charges \cite{R7} correspond to independent helicity 
amplitudes
which describe the chargino production processes for polarized
electrons/positrons with negligible lepton masses. Expressed in terms of 
bilinear charges they are collected in Table 1, including the
transformation properties under P and CP.\\
\begin{table}[\hbt]
\caption{\label{tab:quartic}
{\it The independent quartic charges of the chargino system, the measurement 
     of which determines the chargino mass matrix.}}
\begin{center}
\begin{tabular}{|c|c|l|}\hline
 &  &  \\[-4mm]
{\cal P} & {\cal CP} & { }\hskip 2cm Quartic charges \\\hline \hline
 &  &  \\[-3mm]
 even    &  even     & $Q_1 =\frac{1}{4}\left[|Q_{RR}|^2+|Q_{LL}|^2
                       +|Q_{RL}|^2+|Q_{LR}|^2\right]$ \\[2mm]
         &           & $Q_2 = \frac{1}{2}{\rm Re}\left[Q_{RR}Q^*_{RL}
                       +Q_{LL}Q^*_{LR}\right]$ \\[2mm]
         &           & $Q_3 = \frac{1}{4}\left[|Q_{RR}|^2+|Q_{LL}|^2
                       -|Q_{RL}|^2-|Q_{LR}|^2\right]$ \\[2mm]
         &           & $Q_5=\frac{1}{2}{\rm Re} \left[Q_{LR}Q^*_{RR}
                       +Q_{LL}Q^*_{RL}\right]$ \\
 & & \\[-3mm]
\cline{2-3} 
 & & \\[-3mm]
         &  odd      & $Q_4=\frac{1}{2}{\rm Im}\left[Q_{RR}Q^*_{RL}
                       +Q_{LL}Q^*_{LR}\right]$\\[2mm] \hline \hline
 & & \\[-3mm]
 odd     &  even     & $Q'_1=\frac{1}{4}\left[|Q_{RR}|^2+|Q_{RL}|^2
                        -|Q_{LR}|^2-|Q_{LL}|^2\right]$\\[2mm]
         &           & $Q'_2=\frac{1}{2}{\rm Re}\left[Q_{RR}Q^*_{RL}
                        -Q_{LL}Q^*_{LR}\right]$ \\[2mm]
         &           & $Q'_3=\frac{1}{4}\left[|Q_{RR}|^2+|Q_{LR}|^2
                        -|Q_{RL}|^2-|Q_{LL}|^2\right]$\\[2mm] 
\hline
\end{tabular}
\end{center}
\end{table}

The charges $Q_1$ to $Q_3$, and $Q_5$, $Q_4$ are manifestly parity--even,
$Q'_1$ to $Q'_3$ are parity--odd. The charges 
$Q_1$ to $Q_3$, $Q_5$, and $Q'_1$ to $Q'_3$ are CP invariant\footnote{ 
When expressed in terms of the fundamental SUSY parameters, 
these charges do depend nevertheless on $\cos\Phi_\mu$ indirectly through 
$\cos 2\phi_{L,R}$, in the same way as the $\tilde{\chi}^\pm_{1,2}$
masses depend indirectly on 
this parameter.} 
while $Q_4$ changes sign under CP transformations\footnote{The P--odd and
CP--even/CP-odd counterparts to $Q_5/Q_4$, which carry a negative
sign between the corresponding $L$ and $R$ components, do not affect 
the observables under consideration.}.
The CP invariance of $Q_2$ and $Q'_2$ can easily be proved by noting that
\begin{eqnarray}
2 m_{\tilde\chi_1^\pm} m_{\tilde\chi_2^\pm} 
\cos(\beta_L-\beta_R+\gamma_1-\gamma_2)
\sin2\phi_L\sin2\phi_R \nonumber\\    
= (m^2_{\tilde\chi^\pm_1}+m^2_{\tilde\chi^\pm_2})
      \left(1-\cos2\phi_L \cos2\phi_R\right)
 -4m^2_W 
\label{eq:sisisi} 
\end{eqnarray}
Therefore, all the production cross sections 
$\sigma[e^+e^-\rightarrow \tilde{\chi}^+_i\tilde{\chi}^-_j]$ for any 
combination of pairs $\{ij\}$ depend only on  $\cos 2 \phi_L$
and $\cos 2 \phi_R$ apart from the chargino masses, the sneutrino
mass and the Yukawa couplings. For longitudinally--polarized electron beams 
the sums and differences of the quartic charges are restricted to either 
$L$ or $R$ components (first index) of the $e^\pm$ currents.\\ 

Defining the $\tilde{\chi}^-_i$ production 
angles with respect to the electron flight--direction by the polar angle 
$\Theta$ and the azimuthal angle $\Phi$ with respect to the electron 
transverse polarization, the helicity amplitudes 
can be derived from eq.(\ref{eq:production amplitude}). While electron
and positron helicities are opposite to each other in all amplitudes,
the $\tilde{\chi}^-_i$ and $\tilde{\chi}^+_j$ helicities are in
general not correlated due to the non--zero masses of the particles;
amplitudes with equal $\tilde{\chi}^-_i$ and $\tilde{\chi}^+_j$
helicities are reduced only to order  
$\propto m_{\tilde{\chi}^\pm_{i,j}} /\sqrt{s}$ 
for asymptotic energies. The helicity amplitudes
may be expressed as  
$T_{ij}(\sigma;\lambda_i,\lambda_j)=2\pi\alpha\,{\rm e}^{i\sigma\Phi}\langle
\sigma;\lambda_i\lambda_j\rangle$, 
denoting the electron helicity by the first index, the
$\tilde{\chi}^-_i$ and $\tilde{\chi}^+_j$ helicities by the remaining
two indices, $\lambda_i$ and $\lambda_j$, respectively.
The explicit form of the helicity amplitudes  
$\langle\sigma;\lambda_i\lambda_j\rangle$ can be found in 
Ref.~\cite{CDSZ}.\\ 

In order to describe the electron and positron polarizations, 
the reference frame must be fixed. The electron momentum direction
will define the $z$-axis and the electron transverse polarization vector 
the $x$-axis.
The azimuthal angle of the transverse polarization vector of the positron
is called $\eta$ with respect to the $x$--axis. In this notation, 
the polarized differential cross section 
is given in terms of the electron and positron polarization vectors 
$P$=$(P_T,0,P_L)$ and $\bar{P}$=$(\bar{P}_T \cos\eta,\bar{P}_T\sin\eta,
-\bar{P}_L)$ by
\begin{eqnarray}
\frac{{\rm d}\sigma}{{\rm d}\Omega}
  =\frac{\alpha^2}{16 s} \lambda^{1/2} \bigg[
     (1-P_L\bar{P}_L)\Sigma_{\rm unp}+(P_L-\bar{P}_L)\Sigma_{LL}
  +P_T\bar{P}_T\cos(2\Phi-\eta)\Sigma_{TT}\bigg]\
\end{eqnarray}
with the distributions 
\begin{eqnarray}
\Sigma_{\rm unp}
 &=&\frac{1}{4}\sum_{\lambda_i\lambda_j}
   \left[|\langle +;\lambda_i\lambda_j\rangle|^2
        +|\langle -;\lambda_i\lambda_j\rangle|^2\right] \nonumber\\
\Sigma_{LL}
 &=&\frac{1}{4}\sum_{\lambda_i\lambda_j}
   \left[|\langle +;\lambda_i\lambda_j\rangle|^2
        -|\langle -;\lambda_i\lambda_j\rangle|^2\right] \nonumber\\
\Sigma_{TT}
 &=&\frac{1}{2}\sum_{\lambda_i\lambda_j}
   {\rm Re}\left[\langle -;\lambda_i\lambda_j\rangle
                 \langle +;\lambda_i\lambda_j\rangle^*\right]
\end{eqnarray}
which depend only on the polar angle $\Theta$, but do not on the azimuthal 
angle $\Phi$ any more;
$\lambda=[1-(\mu_i+\mu_j)^2][1-(\mu_i-\mu_j)^2]$ is the two--body 
phase space function and $\mu_i^2=m^2_{\tilde\chi_i^\pm}/s$.\\

Carrying out the sum over the chargino helicities,
the distributions $\Sigma_{\rm unp}$, $\Sigma_{LL}$, and $\Sigma_{TT}$
can be expressed in terms of the quartic charges:
\begin{eqnarray}
\Sigma_{\rm unp}&=& 4\,\bigg\{\left[1-(\mu^2_i - \mu^2_j)^2
                   +\lambda\cos^2\Theta\right]Q_1
                   +4\mu_i\mu_j Q_2+2\lambda^{1/2} Q_3\cos\Theta\bigg\}
                  \nonumber\\
\Sigma_{LL}     &=& 4\,\bigg\{\left[1-(\mu^2_i - \mu^2_j)^2
                   +\lambda\cos^2\Theta\right]Q'_1
                   +4\mu_i\mu_j Q'_2+2\lambda^{1/2} Q'_3\cos\Theta\bigg\}
                  \nonumber\\
\Sigma_{TT}     &=&-4\lambda \sin^2\Theta\,\, Q_5
\end{eqnarray}
If the production angles could be measured unambiguously on an event--by--event
basis, the quartic charges could be extracted directly from the angular 
dependence of the cross section at a single energy. However, since charginos 
decay mainly into the invisible lightest neutralinos and SM fermion pairs, 
the production angles cannot be determined completely
on an event-by-event basis. The transverse distribution can be extracted by 
using an appropriate weight function for the azimuthal angle $\Phi$,
cf. eq.(16). 
This leads us to the 
following integrated polarization--dependent cross sections
as physical observables:
\begin{eqnarray}
\sigma_R&=&\int{\rm d}\Omega\,\,\frac{{\rm d}\sigma}{{\rm d}\Omega}
              \left[P_L=-\bar{P}_L=+1\right] \nonumber\\
\sigma_L&=&\int{\rm d}\Omega\,\,\frac{{\rm d}\sigma}{{\rm d}\Omega}
              \left[P_L=-\bar{P}_L=-1\right] \nonumber\\
\sigma_T&=&\int{\rm d}\Omega\,\,\left(\frac{\cos 2\Phi}{\pi}\right)
               \frac{{\rm d}\sigma}{{\rm d}\Omega}
              \left[P_T=\bar{P}_T=1,\eta=\pi\right]
\label{eq:xsections}
\end{eqnarray}
As a result, nine independent physical observables can be constructed at
a given c.m. energy by
means of beam polarization in the three production processes; three in each 
mode $\{ij\}=\{11\},\{12\}$ and $\{22\}$.

\section*{3.~Measuring Masses, Mixing Angles and Couplings}

Before the strategies to measure the masses, mixing angles and the couplings
are presented in detail, a few general remarks on the structure of the 
chargino system may render the techniques more transparent.\\

\noindent
(i) The right--handed cross sections $\sigma_R$ do not involve the exchange 
    of the sneutrino. They depend only, in symmetric form, on the mixing 
    parameters $\cos 2\phi_L$ and $\cos 2\phi_R$.\\

\noindent
(ii) The left--handed cross sections $\sigma_L$ and the transverse cross
     section $\sigma_T$ depend on $\cos 2\phi_{L,R}$, the sneutrino
     mass and the $e\tilde{\nu}\tilde{W}$ Yukawa coupling. Thus the sneutrino 
     mass and the Yukawa coupling can be determined from the left-handed and 
     transverse cross sections. 
     [If the sneutrino mass is much larger than the collider energy, 
     only the ratio of the Yukawa coupling over the sneutrino mass squared
     ($g^2_Y/m^2_{\tilde\nu}$) can be measured by this method \cite{gmpsnu}.]\\

The cross sections $\sigma_L$, $\sigma_R$ and $\sigma_T$ are binomials
in the [$\cos2\phi_L,\cos2\phi_R$] plane. If the two--chargino model is 
realized in nature, any two $\sigma_L$ and $\sigma_R$ contours, for example, 
will at least cross at one point in the plane between $-1 \leq \cos2\phi_L,
\cos2\phi_R \leq +1$.  However, being ellipses or hyperbolae, they may cross 
up to four times. This ambiguity can be resolved by measuring the third 
physical quantity $\sigma_T$ for example. 
The measurement of $\sigma_T$ is particularly important if the sneutrino
mass is still unknown. While the curve for $\sigma_R$ is fixed, the curve
for $\sigma_L$ will move in the $[\cos2\phi_L,\cos2\phi_R]$ plane with 
changing $m_{\tilde\nu}$. However, the third curve will intersect the other 
two in the same point only if the mixing angles as well as the sneutrino 
mass are chosen right.\\

The numerical analyses presented below  have been worked out for the two 
parameter points introduced in Ref.~\cite{LCWS}.     
They correspond to a
small and a large $\tan\beta$ solution for universal gaugino and
scalar masses at the GUT scale:
\begin{eqnarray}
&&\mbox{\boldmath $RR1$}:
  (\tan\beta, m_0,M_{\frac{1}{2}})=(\,\,3,100\,{\rm GeV}, 200\, {\rm GeV})
  \nonumber\\
&&\mbox{\boldmath $RR2$}:
  (\tan\beta, m_0,M_{\frac{1}{2}})=(30, 160\,{\rm GeV}, 200\, {\rm GeV})
\label{eq:parameter}
\end{eqnarray}
The CP-phase $\Phi_\mu$ is set zero. The induced chargino
$\tilde{\chi}^\pm_{1,2}$, neutralino $\tilde{\chi}^0_1$ and sneutrino
$\tilde{\nu}$ masses are given as follows:
\begin{eqnarray}
&& m_{\tilde{\chi}^\pm_1}=128/132\,{\rm GeV}\,  \qquad
   m_{\tilde{\chi}^0_1}\,  = 70/72\,{\rm GeV}\, \nonumber\\
&& m_{\tilde{\chi}^\pm_2}=346/295\,{\rm GeV}\, \qquad
   m_{\tilde{\nu}}       =166/206 \,{\rm GeV}
\label{eq:measured masses}
\end{eqnarray}
for the two points \mbox{\boldmath $RR1/2$}, respectively.
The size of the unpolarized total cross sections
$\sigma[e^+e^- \rightarrow \tilde\chi_i^+ \tilde\chi_j^-]$ as functions 
of the collider energy is shown for two reference points in Fig.~2. 
With the 
maximum of the cross sections in the range of 0.1 to 0.3~pb, about 
$10^5$ to $3\times10^5$ events can be generated for an integrated luminosity
$\int {\cal L} \simeq 1 {\rm ab}^{-1}$ as planned in three years of running 
at TESLA. The cross sections rise steeply at the threshold,
\begin{eqnarray}
\sigma[e^+e^- \rightarrow \tilde\chi_i^+ \tilde\chi^-_j] \sim
\sqrt{s-(m_{\tilde{\chi}^\pm_i}+m_{\tilde{\chi}^\pm_j})^2}
\end{eqnarray}
so that the masses $m_{\tilde\chi^\pm_1}$, $m_{\tilde\chi^\pm_2}$
can be measured very accurately in the production processes of the
final--state pairs $\{11\}$, $\{12\}$ and $\{22\}$. 
Detailed experimental simulations 
have shown that 
accuracies $\Delta m_{\tilde\chi_1^\pm}$ =~40~MeV and 
$\Delta m_{\tilde\chi_2^\pm}$=250~MeV can be achieved in 
high-luminosity threshold scans \cite{martyn}.

\bigskip

\section*{3.1 Light Chargino Pair Production}

At an early phase of the $e^+e^-$ linear collider the energy may only be
sufficient to reach the threshold of the lightest chargino pair
$\tilde{\chi}^+_1\tilde{\chi}^-_1$. Nevertheless, nearly the entire
structure of the chargino system can be reconstructed even in this case.\\

By analyzing the $\{11\}$ mode in $\sigma_L\{11\}$, $\sigma_R\{11\}$, the
mixing angles $\cos2\phi_L$ and $\cos2\phi_R$ can be determined up to
at most a four fold ambiguity if the sneutrino mass is known 
and the Yukawa coupling is identified with the gauge coupling. 
The ambiguity can be resolved by adding the information from 
$\sigma_T\{11\}$. This is clearly demonstrated\footnote{With event numbers of
order $10^5$, statistical errors are at the per--mille level.} in Fig.~3, 
for the reference point \mbox{\boldmath $RR1$}
at the energy $\sqrt{s}=400$ GeV. 
Moreover, the additional measurement of the
transverse cross section can also be exploited to determine the
sneutrino mass. While the right--handed cross section $\sigma_R$ 
does not depend on $m_{\tilde \nu_e}$, the contours $\sigma_L$, $\sigma_T$ 
move uncorrelated in the $[\cos2\phi_L,\cos2\phi_R]$ plane if not the correct 
sneutrino mass is used in the analysis. The three contour lines intersect 
exactly in one point of the plane only if all the parameters correspond to the
correct physical values.\\

Thus, with transverse polarization available, 
the system of observables can be closed except for the mass value
of the heavy chargino. It has been proved in Ref.~\cite{CDSZ} that
the SUSY parameters $\{M_2,\mu,\tan\beta\}$ can be derived from the
observables $m_{\tilde{\chi}^\pm_1}$ and $\cos 2\phi_{L,R}$ up to at
most a two--fold ambiguity.

\section*{3.2 The Complete Chargino System}

From the analysis of the complete chargino system 
$\{\tilde\chi_1^+ \tilde\chi_1^-, \tilde\chi_1^+ \tilde\chi_2^-,
\tilde\chi_2^+ \tilde\chi_2^-\}$, together with the knowledge of the 
sneutrino mass from sneutrino pair production, the maximal information 
can be extracted on the basic parameters of the electroweak SU(2) gaugino 
sector. Moreover, the identity of the $e \tilde\nu \tilde W$ Yukawa 
coupling with the $e \nu W$ gauge coupling, which is of fundamental nature
in supersymmetric theories, can be tested very accurately. This 
analysis is the final target of LC experiments which should provide
a complete picture of the electroweak gaugino sector with resolution at 
least at the per-cent level.\\

The case will be exemplified for the scenario 
\mbox{\boldmath $RR1$} with $\tan\beta=3$. 
To simplify the picture, without loss of generality, we will not choose 
separate energies at the maximal values of the cross sections, but instead 
we will work at a single collider energy $\sqrt{s}$=
800 GeV and an integrated luminosity $\int {\cal L} =1 {\rm ab}^{-1}$. 
The polarized cross sections take the following values:
\begin{eqnarray}
\begin{array}{lll}
 \sigma_R\{11\}=\,\,1.8\,{\rm fb}\, &{ }\hskip 3mm
 \sigma_L\{11\}=787.7\,{\rm fb}\,   &{ }\hskip 3mm
 \sigma_T\{11\}=0.53\,{\rm fb}  \\
 \sigma_R\{12\}=12.1\,{\rm fb}\, &{ }\hskip 3mm
 \sigma_L\{12\}=106.2\,{\rm fb}\,&{ }\hskip 3mm
 \sigma_T\{12\}=0.53\,{\rm fb} \\
 \sigma_R\{22\}=67.1\,{\rm fb}\, &{ }\hskip 3mm
 \sigma_L\{22\}=337.5\,{\rm fb}\, & { }\hskip 3mm
 \sigma_T\{22\}=1.07\,{\rm fb}
\end{array}
\label{eq:measured}
\end{eqnarray}

Chargino pair production with right-handed electron beams provides us 
with the cross sections $\sigma_{R_i}$ ($i=\{11\},\{12\},\{22\}$). Due to the 
absence of the sneutrino exchange diagram, the cross sections can be 
expressed symmetrically in the mixing parameters 
$c_{2L}=\cos2\phi_L$ and $c_{2R}=\cos2\phi_R$:
\begin{eqnarray}
\sigma_{R_i} = A_{R_i}\,(c^2_{2L}+c^2_{2R})+B_{R_i}\,(c_{2L}+c_{2R})
+C_{R_i}\, c_{2L}c_{2R}+D_{R_i}\ \ \left(i=\{11\},\{12\},\{22\}\right)
\label{eq:sigmar}
\end{eqnarray}
The coefficients $A_{R_i}$, $B_{R_i}$, $C_{R_i}$ and $D_{R_i}$  
involve only known parameters, the chargino masses and the
energy. Depending on
whether $A^2_{R_i} \stackrel{>}{{ }_<} C^2_{R_i}/4$, the contour
lines in the [$c_{2L},c_{2R}$] plane (cf. Fig.4) are either 
closed ellipses or open hyperbolae\footnote{The cross
section $\sigma_R\{12\}$ is always represented by an ellipse.}. 
They intersect in exactly two 
points in the plane which are symmetric under the interchange $c_{2L}
\leftrightarrow c_{2R}$; for \mbox{\boldmath $RR1$}: [$c_{2L},c_{2R}$]
=[0.645,0.844]
and interchanged.\\

While the right--handed cross sections do not involve sneutrino exchange,
the cross sections for left-handed electron beams are dominated 
by the sneutrino contributions unless the sneutrino mass is very large. 
In general, the three observables $\sigma_{L_i}$ ($i=\{11\},\{12\},\{22\}$) 
exhibit quite a different dependence on $c_{2L}$ and  
$c_{2R}$. In particular, they are not symmetric with respect 
to $c_{2L}$ and $c_{2R}$ so that   
the correct solution for $[c_{2L},c_{2R}]$ can be 
singled out of the two solutions obtained from the right-handed cross 
sections eq.(\ref{eq:sigmar}). As before, the three observables 
can be expressed as   
\begin{eqnarray}
\sigma_{L_i}= A_{L_i}\,c^2_{2L}+A'_{L_i}\,c^2_{2R}+ B_{L_i}\,c_{2_L}
           +B'_{L_i}\,c_{2R}+C_{L_i}\,c_{2L}c_{2R}+D_{L_i}\,  
	   \left(i=\{11\},\{12\},\{22\}\right)
\label{eq:sigmal}
\end{eqnarray}
The coefficients of the linear and quadratic terms of $c_{2L}$ and 
$c_{2R}$ depend on known parameters only. The shape of the contour lines
is given by the chargino masses and the sneutrino mass, being either elliptic 
or hyperbolic for $A_{L_i} A'_{L_i} \stackrel{>}{{ }_<} C^2_{L_i}/4$, 
respectively. These asymmetric equations are satisfied {\it only} by one 
solution, as shown in Fig.~4. Among the two 
solutions obtained above from $\sigma_{R_i}$ only the set 
$[c_{2L},c_{2R}]=[0.645,0.844]$ 
satisfies eq.(\ref{eq:sigmal}).

At the same time, the identity between the $e \tilde\nu \tilde W$
Yukawa coupling and the $e \nu W$ gauge coupling can be tested.
\hskip -2mm Varying 
the Yukawa coupling freely, the contour lines $\sigma_{L_i}$ are shifted
through the [$c_{2L},c_{2R}$] plane. Only for the supersymmetric 
solutions the curves $\sigma_{L_i}$ intersect each other and the curves
$\sigma_{R_i}$ in exactly one point. Combining the analyses of 
$\sigma_{R_i}$ and $\sigma_{L_i}$, the masses, the mixing
parameters and the Yukawa coupling can be determined to quite a high 
precision\footnote{In contrast to the restricted
$\tilde{\chi}^+_1\tilde{\chi}^-_1$ case, it is not 
necessary to use transversely polarized beams to determine this set of 
parameters unambiguously. If done so nevertheless, the analysis follows the 
same steps as discussed above. The additional information will reduce the 
errors on the fundamental parameters.}
\begin{eqnarray}
&&{ }\hskip -1.2cm m_{\tilde\chi_1^\pm}=128\pm 0.04\, {\rm GeV}\quad 
   \cos2\phi_L=0.645 \pm 0.02  \qquad
   g[e\tilde\nu \tilde W]/g[e\nu W]= 1\pm 0.001 \nonumber\\  
&&{ }\hskip -1.2cm m_{\tilde\chi_2^\pm}=346\pm 0.25\, {\rm GeV}\quad
   \cos2\phi_R=0.844 \pm 0.005 
\label{eq:measured}
\end{eqnarray}
The 1$\sigma$ level statistical errors have been derived for an integrated 
luminosity of $\int {\cal L} =1\,{\rm ab}^{-1}$.\\

Thus the parameters of the chargino system, masses $m_{\tilde\chi_1^\pm}$
and  $m_{\tilde\chi_2^\pm}$, mixing parameters $\cos2\phi_L$ and 
$\cos2\phi_L$, as well as the Yukawa coupling can be used to extract the 
fundamental parameters of the underlying supersymmetric theory
with high accuracy.

\section*{\bf 4. Deriving the Fundamental SUSY Parameters}

From the set (\ref{eq:measured}) of measured observables the fundamental 
supersymmetric parameters $\{M_2, \mu, \cos\phi_\mu, \tan\beta\}$ can be
derived in the following way. To compactify the expressions, we introduce
the abbreviations
\begin{eqnarray}
&& \Sigma =(m^2_{\tilde{\chi}^\pm_2}+m^2_{\tilde{\chi}^\pm_1}-2m^2_W)/2m^2_W
   \nonumber\\
&& \Delta=(m^2_{\tilde{\chi}^\pm_2}-m^2_{\tilde{\chi}^\pm_1})/4m^2_W
\end{eqnarray}
where $\Delta$ is defined equivalently to eq.(\ref{eq:delta}).\\

\noindent
{\bf (i) \boldmath{$M_2,|\mu|$}} -- Based on  the definition 
$M_2>0$, the gaugino mass parameter $M_2$ and the modulus of the higgsino
mass parameter read as follows:
\begin{eqnarray}
  M_2&=&m_W\sqrt{\Sigma-\Delta(c_{2L}+c_{2R})}\nonumber\\
|\mu|&=&m_W\sqrt{\Sigma+\Delta (c_{2L}+c_{2R})}
\label{eq:M2mu}
\end{eqnarray} 

\mbox{ } \\[-4mm]

\noindent
{\bf (ii) $\mbox{\boldmath{$\cos\Phi$}}_\mu$} --
The sign of $\mu$ in CP--invariant theories and, more generally, the cosine 
of the phase of $\mu$ in CP--noninvariant theories is determined by 
the $\tilde{\chi}^\pm_1,\tilde{\chi}^\pm_2$ masses and $\cos 2\phi_{L,R}$:
%
\begin{eqnarray}
\cos\Phi_\mu=\frac{\Delta^2 (2-c^2_{2L}-c^2_{2R})-\Sigma }{
                   \sqrt{[1-\Delta^2 (c_{2L}-c_{2R})^2]
                         [\Sigma^2 -\Delta^2(c_{2L}+c_{2R})^2]}}\,
\label{eq:phi_mu}
\end{eqnarray}
\noindent
{\bf (iii) \boldmath{$\tan\beta$}} -- The value of $\tan\beta$ 
is uniquely determined in terms of two chargino masses and two mixing 
angles
\begin{eqnarray}
\tan\beta=\sqrt{\frac{1-\Delta (c_{2L}-c_{2R})}{
                      1+\Delta (c_{2L}-c_{2R})}}\,
\label{eq:tanb}
\end{eqnarray}
As a result, the fundamental SUSY parameters $\{
M_2,\mu, \tan\beta\}$ in CP--invariant
theories, and $\{M_2, |\mu|, \cos\Phi_\mu, \tan\beta\}$
in CP--noninvariant theories, 
can be extracted {\rm unambiguously} from the observables 
$m_{\tilde{\chi}^\pm_{1,2}}$, $\cos 2\phi_R$, and $\cos 2\phi_L$.
The final ambiguity in $\Phi_\mu \leftrightarrow 2 \pi - \Phi_\mu$ 
in CP--noninvariant theories must be resolved by measuring observables
related to the 
normal $\tilde{\chi}^-_1$ or/and $\tilde{\chi}^+_2$  polarization in 
non--diagonal $\tilde{\chi}^-_1\tilde{\chi}^+_2$ chargino--pair 
production \cite{R11}.\\
 
For illustration, the accuracy which can be expected in such an analysis,
is shown for both CP invariant reference points \mbox{\boldmath $RR1$} 
and \mbox{\boldmath $RR2$} 
in Table 2.  If $\tan\beta$ is large, this parameter is difficult to
extract from the chargino sector. Since the chargino observables depend 
only on $\cos2\beta$, the dependence on $\beta$ is flat for $\beta\rightarrow 
\pi/2$ so that eq.(\ref{eq:tanb}) is not very useful to derive the value of
$\tan\beta$ due to error propagation. A significant lower bound can
be derived nevertheless in any case. 

%
\begin{table}[\hbt]
\caption{\label{tab:measured}
{\it Estimate of the accuracy with which the parameters $M_2$, $\mu$,
     $\tan\beta$ can be determined, including sgn($\mu$),
     from chargino masses and production cross sections; errors are
     statistical only at the 1$\sigma$ level.}}
\begin{center}
\begin{tabular}{|c|c|c|c|c|}\hline
    & \multicolumn{2}{|c|}{\mbox{\boldmath $RR1$}} & 
      \multicolumn{2}{c|}{\mbox{\boldmath $RR2$}}\\
  \cline{2-5}
& theor. value & fit value & theor. value & fit value \\ 
  \cline{2-5} \hline 
  &  &  &  & \\[-3mm] 
$M_2$      & 152\, GeV & $152\pm 1.75$\, GeV & 150\, GeV & $150\pm 1.2$\, GeV \\
$\mu$      & 316\, GeV & $316\pm 0.87$\, GeV & 263\, GeV & $263\pm 0.7$\, GeV \\
$\tan\beta$& 3         & $3\pm 0.69$         & 30        &  $> 20.2$   \\[1mm]
\hline
\end{tabular}
\end{center}
\end{table}
%


\section*{5. Sum Rules} 

The two--state mixing of charginos leads to sum rules for the chargino
couplings. They can be formulated in terms of the squares of the
bilinear charges, {\it i.e.} the elements of the quartic charges.
This follows from the observation that the mixing matrix is
built up by trigonometric functions among which many relations are
valid.  From evaluating these sum rules experimentally, it can be concluded 
whether $\{\tilde\chi_1^\pm, \tilde\chi_2^\pm \}$ forms a closed system,
or whether additional states, at high mass scales, mix in.\\

The following general sum rules can be derived for the two--state 
charginos system at tree level:
\begin{eqnarray}
\sum_{i,j=1,2}|Q_{\alpha\beta}|^2\{ij\} 
  =
2\left(|D_\alpha|^2+|F_\alpha|^2\right) \quad\qquad (\alpha\beta)=(LL,RL,RR) 
\label{eq:sum rule}
\end{eqnarray}
The right--hand side is independent of any supersymmetric parameters, and
it depends only on the electroweak parameters $\sin^2\theta_W, m_Z$ 
and on the energy, cf. eq.(\ref{eq:DFLR}).  Asymptotically, the initial 
energy dependence and the $m_Z$ dependence drop out.
The corresponding sum rule for the mixed left--right (LR) combination,
\begin{eqnarray}
\sum_{i,j=1,2}|Q_{LR}|^2\{ij\}=2\left(|D'_L|^2+|F'_L|^2\right)
\end{eqnarray}
%
involves the sneutrino mass and Yukawa coupling.\\
  
The validity of these sum rules is reflected in both the quartic 
charges and the production cross sections. However, due to mass effects
and the $t$-channel sneutrino exchange, it is not straightforward to derive 
the sum rules for the quartic charges and the production cross sections 
in practice. Only {\it asymptotically} at high energies the sum rules 
(\ref{eq:sum rule})
for the charges can be transformed directly into sum rules for the
associated cross sections.
Nevertheless, the fact that all the physical 
observables are bilinear in $\cos2\phi_L$ and $\cos2\phi_R$, enables 
us to relate the cross
sections with the set of the six variables 
$\vec{z}=\{1,c_{2L},c_{2R},c^2_{2L}, c^2_{2R},c_{2L}c_{2R}\}$. 
For the sake of simplicity we restrict ourselves to the left and 
right--handed cross section. We introduce the
generic notation $\vec{\sigma}$ for the
six cross sections $\sigma_R\{ij\}$ and $\sigma_L\{ij\}$:
\begin{eqnarray}
\vec{\sigma} =\bigg[\,\sigma_R\{11\},\sigma_R\{12\},\sigma_R\{22\},
                     \sigma_L\{11\},\sigma_L\{12\},\sigma_L\{22\}\,\bigg]\,
\end{eqnarray}
Each cross section can be decomposed in terms of $c_{2L}$ and $c_{2R}$ as
\begin{eqnarray}
{\sigma}_i = \sum_{j=1}^6 \, 
    f_{ij}[m^2_{\tilde{\chi}^\pm_{1,2}},m^2_{\tilde{\nu}}] \,z_j
\end{eqnarray}
The matrix elements $f_{ij}$ can easily be derived from Table 1 together 
with eqs.(\ref{eq:DFLR}-\ref{eq:[12]}).
Since the observables $\sigma_R$ 
do not involve sneutrino contributions, the corresponding functions 
$f_{ij}$ do not depend on the sneutrino mass.
The 6$\times$6 matrix $f_{ij}$ relates the six left/right-handed cross section
and the six variables $z_i$. Inverting the matrix gives the expressions 
for the variables $z_i$ in
terms of the observables $\sigma_i$, which are not independent. We therefore 
obtain several non--trivial relations among the observables
of the chargino sector: 
\begin{eqnarray}
&& z_1=1\ \ \ \  :\ \ f^{-1}_{1j} {\sigma}_j = 1 \\[1mm]
&& z_4=z^2_2 \, \ \ :\ \ f^{-1}_{4j} {\sigma}_j 
         =\left[ f^{-1}_{2j} {\sigma}_j \right]^2 \\[1mm]
&& z_5=z^2_3\, \ \ : \ \ f^{-1}_{5j} {\sigma}_j 
         =\left[ f^{-1}_{3j} {\sigma}_j\right]^2 \\[1mm]
&& z_6=z_2 z_3 :\ \ f^{-1}_{6j} {\sigma}_j 
         =f^{-1}_{2j} f^{-1}_{3k} {\sigma}_j{\sigma}_k\,
\end{eqnarray}
where summing over repeated indices is understood. The failure of
saturating any of these sum rules by the measured cross sections would
signal that the chargino two--state $\{\tilde\chi_1^\pm$,
$\tilde\chi_2^\pm\}$ system is not complete and additional states mix
in.

\section*{6.~Conclusions}

We have analyzed in this report how the parameters of the chargino 
system, the chargino masses $m_{\tilde{\chi}^\pm_{1,2}}$
and the size of the wino and higgsino components in the chargino 
wave--functions, parameterized by the two mixing angles 
$\phi_L$ and $\phi_R$, can be extracted from pair production of 
the chargino states in $e^+e^-$ annihilation. 
Three production cross sections $\tilde\chi_1 \tilde\chi_1$, 
$\tilde\chi_1 \tilde\chi_2$, $\tilde\chi_2 \tilde\chi_2$,
for left-- and right--handedly polarized electrons
give rise to six independent observables.
The method is independent of the chargino decay properties, {\it i.e.}
the analysis is not affected by the structure of the neutralino sector
which is very complex in supersymmetric theories while the chargino sector 
remains generally isomorphic to the minimal form of the
MSSM.  \\

The measured
chargino masses $m_{\tilde{\chi}^\pm_{1,2}}$ and the 
two mixing angles $\phi_L$ and $\phi_R$ allow us to extract the fundamental 
SUSY parameters $\{M_2,\mu, \tan\beta\}$ in CP--invariant theories
unambiguously; in CP--noninvariant theories the
modulus of $\mu$ and the cosine of the phase can be determined,
leaving us with just a discrete two--fold ambiguity $\phi_\mu \leftrightarrow
2\pi-\phi_\mu$ which can be
resolved by measuring the sign of observables related to the normal 
$\tilde{\chi}^\pm_{1,2}$ polarizations.\\

Sum rules for the production cross sections can be used at high energies
to check whether the two--state chargino system is a closed system or whether
additional states mix in from high scales.\\

{\it To summarize}, the measurement of the processes
$e^+e^-\rightarrow \tilde{\chi}^+_i \tilde{\chi}^-_j$ [$i,j=1,2$] 
carried out with polarized beams, leads to a complete analysis of the basic
SUSY parameters $\{M_2, \mu, \tan\beta\}$ in the chargino sector.
Since the analysis can be performed with high precision, this set provides
a solid platform for extrapolations to scales eventually near the Planck scale
where the fundamental supersymmetric theory may be defined.

\vskip 0.3cm

\section*{Acknowledgments}

This work was supported by the Korea Science and Engineering Foundation
(KOSEF) through the KOSEF--DFG Large Collaboration Project, Project No.
96--0702--01--01--2. MG was supported by the Alexander von Humboldt 
Stiftung.  JK was supported by the KBN grant No. 2P03B 030 14.

\begin{figure}
 \begin{center}
\epsfig{figure=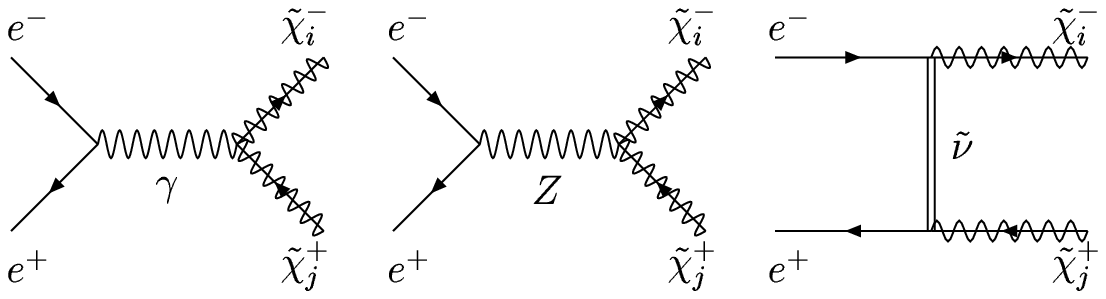,width=16cm,height=7cm}
 \end{center}
\vskip -1cm
\caption{\it The three exchange mechanisms contributing to the production
         of chargino  pairs $\tilde{\chi}^-_i \tilde{\chi}^+_j$ in 
         $e^+e^-$ annihilation.}
\label{fig1}
\end{figure}

\vskip 6cm
\begin{figure}
 \begin{center}
\hbox
to\textwidth{\hss\epsfig{file=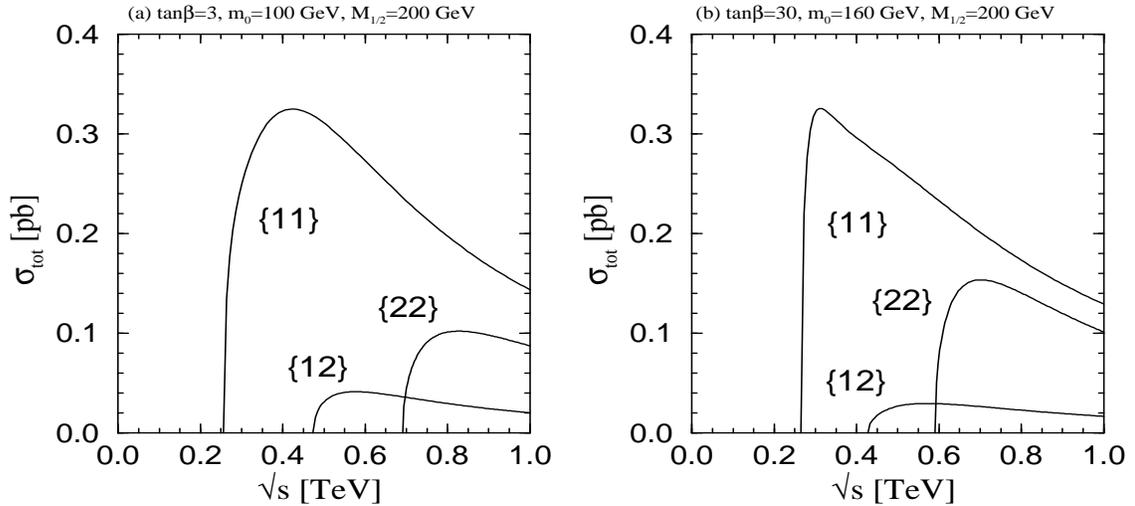,width=15cm,height=7cm}\hss}
 \end{center}
\caption{\it The cross sections for the production of charginos as a
         function of
         the c.m. energy (a) with the $\mbox{\boldmath $RR1$}$ set and
         (b) with the $\mbox{\boldmath $RR2$}$ set of the fundamental
         SUSY parameters.}
\label{xrs}
\end{figure}

\vskip 5cm

\voffset -1cm
\newpage
\vskip -0.6cm
\begin{figure}
 \begin{center}
\hbox to\textwidth{\hss\epsfig{file=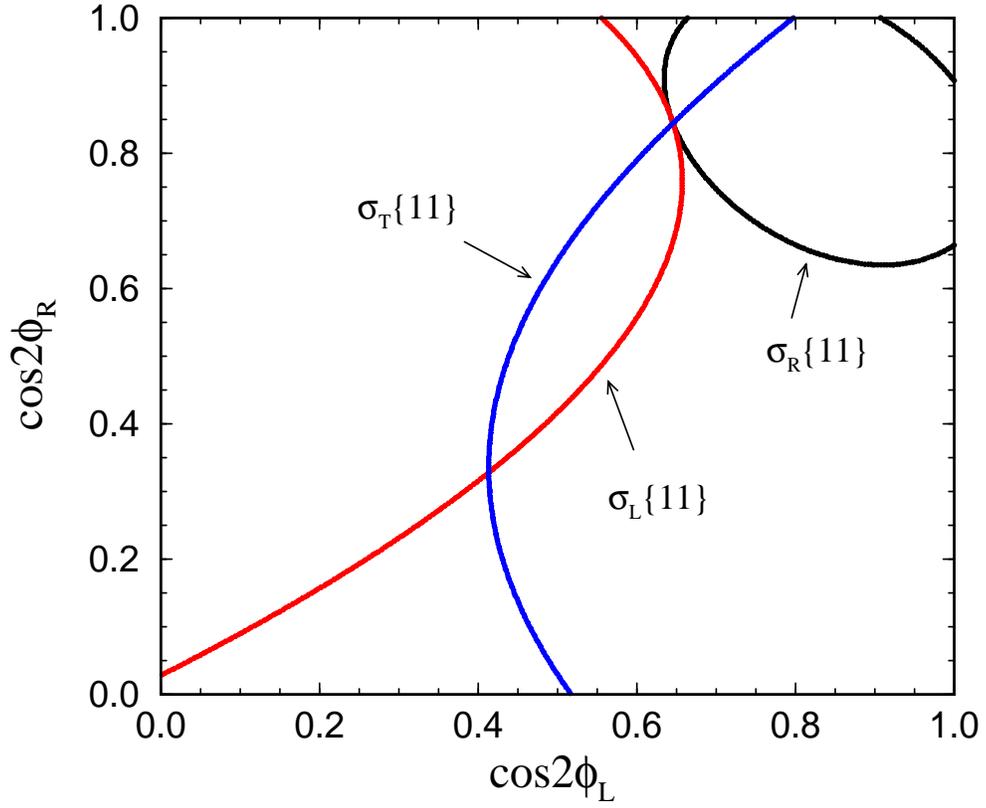,width=13cm,height=11cm}\hss}
 \end{center}
\caption{\it Contours of the cross sections $\sigma_L\{11\}$,~$\sigma_R\{11\}$ 
         and $\sigma_T\{11\}$ in the $[\cos2\phi_L,\cos2\phi_R]$ plane 
	 for the set \mbox{\boldmath $RR1$}
	 $[\tan\beta=3,~m_0=100~{\rm GeV},
         ~M_{1/2}=200~{\rm GeV}]$ at the $e^+e^-$ c.m. energy of 400 GeV.} 
\label{fig3}
\end{figure}

\newpage
\vskip -0.6cm
\begin{figure}
 \begin{center}
\hbox to\textwidth{\hss\epsfig{file=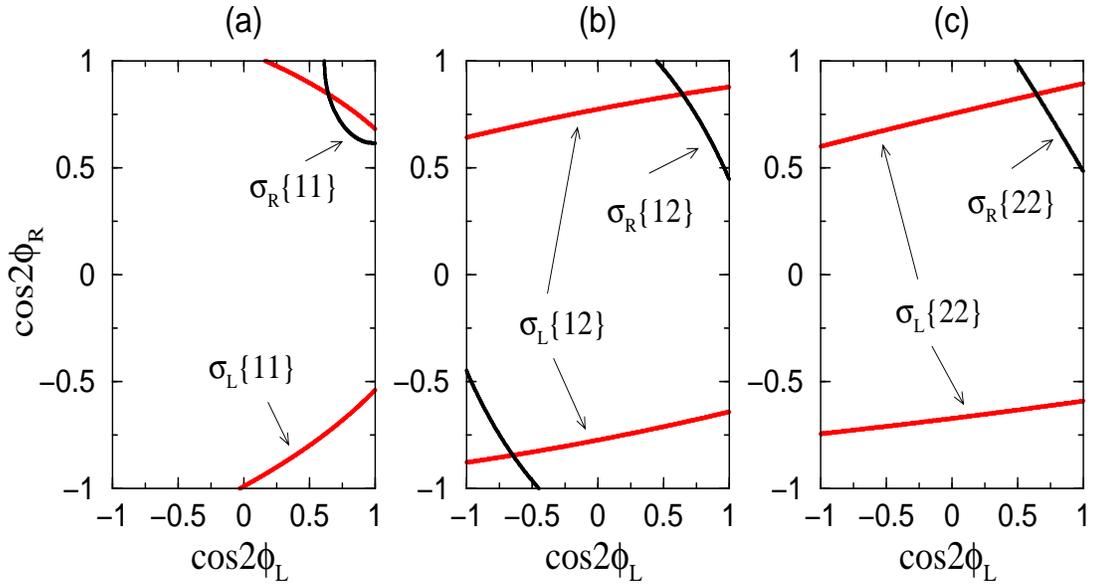,width=14.5cm,height=8cm}\hss}
 \end{center}
\caption{\it Contours of the cross sections 
         (a) $(\sigma_R\{11\},\sigma_L\{11\})$,
	 (b) $(\sigma_R\{12\},\sigma_L\{12\})$ and 
	 (c) $(\sigma_R\{22\},\sigma_L\{22\})$ 
         in the $[\cos2\phi_L,\cos2\phi_R]$ plane 
	 for the set 
	 \mbox{\boldmath $RR1$} $[\tan\beta=3,~m_0=100~{\rm GeV},
         ~M_{1/2}=200~{\rm GeV}]$ at the c.m. energy of 800 GeV.}
\label{fig4}
\end{figure}

\vfil\eject

\end{document}